# New Rain Rate Statistics for Emerging Regions: Implications for Wireless Backhaul Planning

Paul M. Aoki, *Senior Member, IEEE*

*Abstract*— As demand for broadband service increases in emerging regions, high-capacity wireless links can accelerate and cost-reduce the deployment of new networks (both backhaul and customer site connection). Such links are increasingly common in developed countries, but their reliability in emerging regions is questioned where very heavy tropical rain is present. Here, we investigate the robustness of the standard (ITU-R P.837-6) method for estimating rain rates using an expanded test dataset. We illustrate how bias/variance issues cause problematic predictions at higher rain rates. We confirm (by construction) that an improved rainfall climatology can largely address these prediction issues without compromising standard ITU fit evaluation metrics.

## I. Introduction

THE availability and affordability of Internet access is a foundational concern in the field of ICT and development (ICTD) [1] and a key concern for international development as a whole [2]. While prices for network capacity in international transit markets are generally falling due to increased competition, the cost of bringing broadband capacity over land from Internet service provider (ISP) points of presence (PoPs) to cell towers and to end-users remains a key obstacle. This is especially true in emerging regions. Although build costs can be spread across multiple ISPs through the use of a wholesale, or shared infrastructure [3], model, building terrestrial wired networks remains quite expensive – $10/m is optimistic. The business issue here is that infrastructure costs that are considered to be economically sustainable in developed countries may not be sustainable in emerging regions because the customer revenue generated by the corresponding amount of network infrastructure is often lower.

Wireless point-to-point and point-to-multipoint links, especially using unlicensed or lightly-licensed spectrum [4], are often proposed to ameliorate the financial and regulatory issues associated with wired backhaul [5]. In particular, the GHz-scale bandwidths available in millimeter-wave (30-300 GHz) spectrum are becoming increasingly attractive as prices drop (e.g., 1 Gbps radios operating in unlicensed V band spectrum can already be found at costs below $1,000/link).

While low-cost, high-capacity wireless links are already becoming popular to connect urban small cell sites in developed countries [6], their applicability for backhaul in emerging regions – where the need for cost-reduction is greater – is clouded by the issue of rain attenuation and its potential impact on link availability. Locations that frequently experience high rain rates require reduced link distances to protect service levels [7], and increased build density costs more, negatively affecting the business case. (Conversely, showing that locations do not experience frequent high rain rates improves the business case.)

In this paper, we address the question: *where, in emerging regions, are large populations frequently subject to high rain rates?* We care about accuracy as both types of error have business impact and therefore development impact (unexpected increase in build cost for false negatives; opportunity costs for false positives). The research contributions of this paper are as follows:

1. We illustrate, for the first time, the large differences in affected population that result from using different reference datasets (Section III). For example, using various datasets that are widely used in practice and in research, the corresponding estimates of the total population affected by high rain rate in emerging regions varies by 4x and ranges to over a billion people.
2. We construct a novel rainfall dataset for (sub-)tropical latitudes by integrating TRMM satellite radar estimates of near-surface rain [8] on a sub-pixel-resolution grid, improving spatial registration (though not pixel resolution) over prior work (Section IV).
3. We construct an expanded rain rate test set that includes 10x more sites and 2.5x more countries than the current ITU test set [9] in (sub-)tropical latitudes (Section V), and use this expanded test set to assess the accuracy of various methods of estimating rain rate statistics (Section VI).

The somewhat counter-intuitive outcome is that a rain rate estimation method (such as ITU-R P.837-6 [10]) with state-of-the-art performance on the usual fit metrics (e.g., rms error [11]) still has considerable difficulty in accurately identifying which markets (here, countries) are affected by high rain rates. Specifically, because rms error is dominated by its variance component for all methods and all cases analyzed here, methods with comparable rms performance but with a better spatial distribution of bias can be more accurate at this identification task. Our most effective method (so far) results in an estimate of affected population that is ~2.5x greater than the current ITU method. While methods can always be improved (and caveats are discussed in Section VII), we

P. M. Aoki is with Google Inc., 1600 Amphitheatre Parkway, Mountain View, CA, 94043 USA (e-mail: aoki@acm.org).

believe that future revision of ITU-R P.837 is merited and that business case uncertainty of deploying millimeter-wave backhaul in emerging regions may be less than often assumed.

## II. PROBLEM STATEMENT

As we have said, this paper focuses on identifying where large populations are affected by high rain rate in emerging regions.

*Large populations:* All population estimates here are based on the 2012 release of LandScan, a well-known gridded population count product [12]. Population counts are measured infrequently and are not always reliable – executing a census well is difficult and expensive. Gridded population statistics, use models to adjust census totals and "distribute" the population at the desired grid resolution (LandScan's grid resolution is 3 arc-second, or 1/120°).

*High rain rate:* For practical reasons that will we expand upon later, we will focus here on regions where rain rates exceed those characterized in ITU-R PN.837-1 [13] as "Zone N" (e.g., 0.01% exceedance probability rain rate is over 95 mm/h). While PN.837-1 has long been superceded, Zones P and Q are still acknowledged by practicing radio engineers as canonical examples of regions where severe rain fade occurs – Nigeria, Indonesia, Malaysia, etc.

*Emerging regions:* Most of the countries that have, at various times, been labeled "less-developed," "developing," or "emerging" are concentrated in the sub-tropical and tropical latitude band (38°S-38°N). As the main satellite precipitation data that we intend to use is reliable in the latitude range 35°S-35°N, it should be understood throughout the paper that all climate and population datasets (as well as any reported means and totals) are subset to 35°S-35°N (hereafter, the *study area*).

In summary, then, our goal will be to provide gridded rain statistics for 35°S-35°N that not only do well on conventional fit metrics (at least as well as the current ITU statistics) but are demonstrably better at identifying regions exceeding Zone N conditions. The greatest challenge is doing so in a way that, given the shortage of validation data, can be shown to be consistent (at each step) with current understanding of global precipitation.

In this paper, we take the attenuation models themselves as given, and focus entirely on the local rain rate statistics.

## III. BACKGROUND AND MOTIVATION

Obviously, multi-year rain attenuation statistics are not available for all places on Earth. The standard methods of estimating rain attenuation therefore draw on long-term weather datasets to (1) estimate local rain rate statistics and then (2) estimate rain attenuation as a function of proposed link characteristics and the local rain rate statistics [14]. For example, using the ITU-R recommendation for terrestrial wireless links (P.530 [15]), the rain attenuation (in dB/km) that is expected to be exceeded $p$% of the time is calculated as a function of $R_p$, the rain rate (in mm/h) at the corresponding exceedance probability. The rain rates that result in deep rain fades may only be sustained for seconds or minutes, so these rain rates are calculated using a 1-min integration period (hereafter, *1-min rain rate*). By convention [16], the 1-min rain rate statistics for a given site are presented as lists of $R_p$ values at standard exceedance probabilities $p \in \{0.001, 0.002, 0.003, 0.005, \ldots, 1, 2, 3, 5\}$ (hereafter, we call such lists *site statistics*).

### A. Estimating rain rate $R_p$ from rainfall climatologies

The description above seems straightforward, but 1-min rain rate statistics are not widely available either (and almost never on a gratis basis even where they are available). As a result, another level of indirection is usually required.

ITU-R has produced a series of recommendations (P.837) to estimate $R_p$. The first revision, PN.837-1, relied on maps of *rain zones* [16] (drawn by experts [14]) that corresponded to 15 profiles listing $R_p$ at standard exceedance probabilities $p \in \{0.001, 0.003, 0.01, \ldots, 1\}$. Subsequent revisions show how to compute 1-min rain rate from gridded climatological products produced from data collected by operational meteorological networks [17]. For example, P.837-5 [18] takes gridded estimates of three rain statistics (mean annual rainfall $M_t$ in mm, annual probability of rain $P_0$ in %, and convective/stratiform rain fraction $\beta$) derived from the ECMWF ERA-40 reanalysis [19] and plugs them into a fitted model to estimate $R_p$. With interpolation, we can estimate $R_p$ for arbitrary $p$ at any location. To simplify a bit, the model is based on the expression:

$$P(R) = P_0 \exp(-aR(1+bR)/(1+cR)) \quad (1)$$
$$a = x; \; b = M_t/(y \cdot P_0); \; c = z \cdot b$$

solved for $R$; $P_0$ and $M_t$ are the rain statistics at the location in question and the model parameters $x, y, z$ are globally optimized using a non-linear least-squares fit to a hand-picked "training set" of just 31 site statistics [9]. (The current P.837-6 [20] uses the same model.)

The P.837-5/6 model is a remarkable achievement. Although it assumes that the frequency distribution of rain rate at any location can be described using this single family of curves and the fit is based on very limited training/test data, the model produces consistent results (rms error on the order of 30% [9]) over wildly variable global rain conditions and nearly four decades of exceedance probabilities (5% to 0.001%). Its success is suggested by the fact that the model has not been modified since 2007.

TABLE I. GLOBAL ANNUAL RAINFALL CLIMATOLOGIES DISCUSSED IN THIS PAPER (ORDERED BY $\bar{M}_t$).

| short name | description | references | pixel size (°) | source S | source G | $\bar{M}_t$ (mm) land | $\bar{M}_t$ (mm) pop>0 |
|---|---|---|---|---|---|---|---|
| TCC | TRMM Composite Climatology, v2 | [21] | 1/2 | ✓ | | 910 | 1079 |
| WCLIM | WorldClim | [22] | 1/120 | | ✓ | 937 | 1123 |
| GPCC | GPCC Normals v2015 | [23] | 1/4 | | ✓ | 976 | 1167 |
| ITU | ITU-R P.837-5/6 | [10], [18] | 9/8 | ✓ | ✓ | 999 | 1183 |
| TMPA | TRMM Multi-satellite Precipitation Analysis 3B42 (1998-2014) | [24] | 1/4 | ✓ | ✓ | 1014 | 1199 |

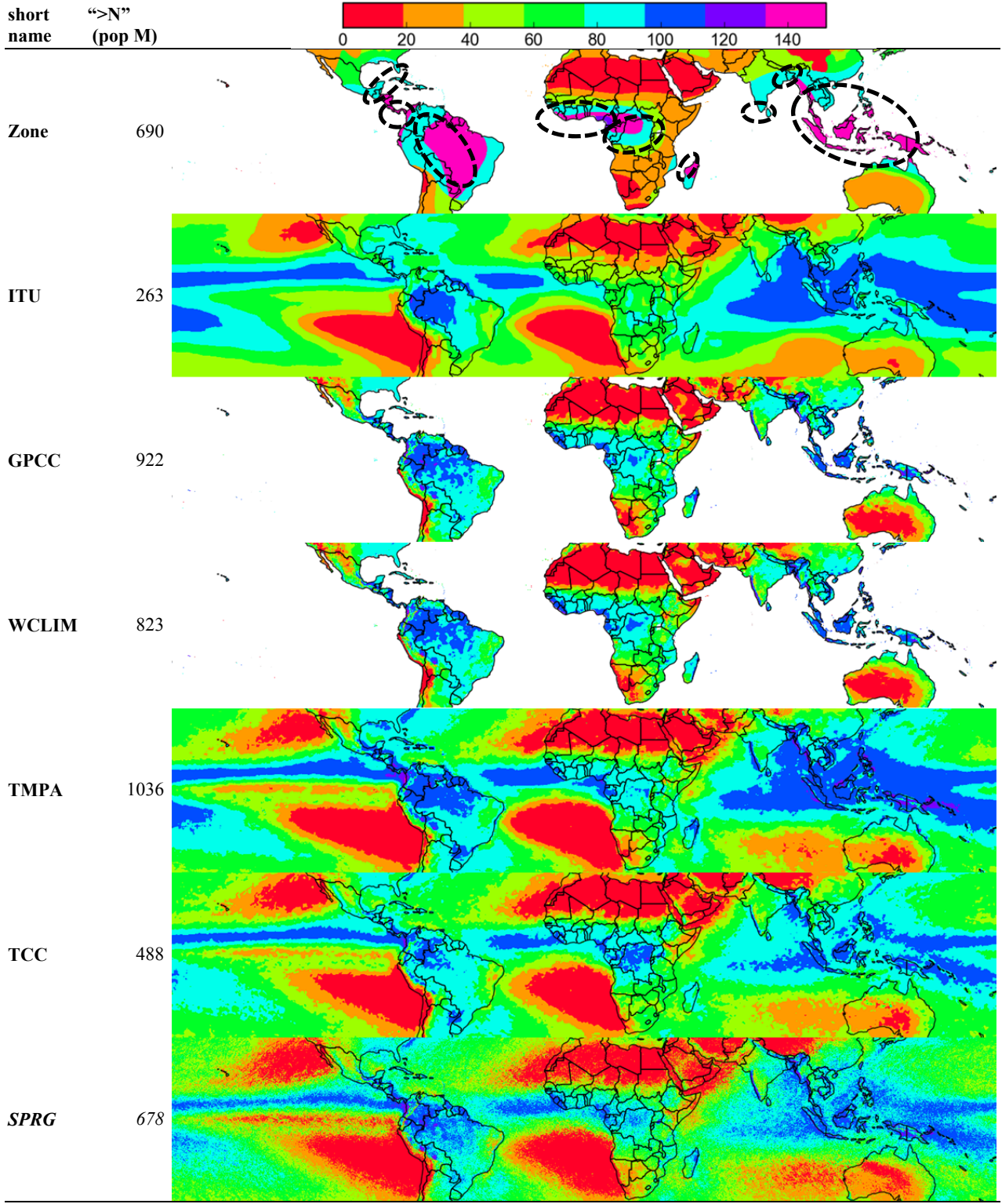

FIGURE 1. RAIN RATE AT 0.01% ANNUAL EXCEEDANCE PROBABILITY $R_{0.01}$ (MM/H, 19 MM/H BINS).
Blue, purple, and magenta pixels all indicate areas that are ">N" (i.e., $R_{0.01}$ >95 mm/h).

*B. Considerable differences exist between climatologies*

Is the ITU method the ultimate solution? There are many rainfall climatologies that, while broadly consistent on a global scale, exhibit considerable local differences [25]. Typically, rainfall climatologies are constructed from the sparse point-measurement data (e.g., rain gauges) that is available, augmented in many cases by indirect area-measurement data (e.g., ground radar, optical observations from high orbit, active and passive microwave observations from low orbit). Source datasets must be merged (in some cases, by assimilation into a climate simulation); the result interpolated onto a grid; and bias (anomaly relative to reference data) corrected. As such, differing choices of source data and algorithms result in variation between climatologies.

Table I describes five notable rainfall climatologies. Of these, two have previously been evaluated for use in this application (ITU [9] and TMPA [26]). We will use the others (TCC, GPCC, WCLIM) as "plug-in" replacements for TMPA to illustrate differences in how these well-known datasets produce different end-results.[1] We focus on the differences resulting from changing $M_t$ as opposed to $P_0$ as it has been noted that the ITU rain rate estimate is more sensitive to $M_t$ than to $P_0$[27].

From the $M_t$ values themselves, it is not clear how important the differences are. Table I shows how (pixel-wise) mean annual rainfall estimates differ. ("Land" means that $M_t$ is averaged over all land pixels in the study area, whereas "pop>0" means that the average is taken over land pixels with non-zero population only.) The differences do not appear to be huge, and there is no comprehensive "ground truth" by which one can validate them [25]. This begs the question: does choice of climatology have an actual impact on $R_p$, and if so, what is its magnitude?

We will focus the discussion here solely on $R_{0.01}$, as many attenuation-related recommendations rely on $R_{0.01}$ [14] (often solely – see, e.g., [15], [28]). Further, we will focus on higher rain rates ($R_{0.01}$> 95 mm/h). The reason to do so in the context of emerging regions is that for RF links above ~20 GHz, a 100 mm/h storm event will cause 10-30 dB/km of rain attenuation loss.[2] Even short (< 1 km) links can be expected to suffer fades or outage under these conditions. As the 0.01% rate for Zone N is 95 mm/h, we will call anything heavier "heavy rain" and designate these regions of interest (where $R_{0.01}$> 95 mm/h) as ">N".

The maps in Figure 1 show $R_{0.01}$ for the rain zone map and the five climatologies of Table I ($R_{0.01}$ always being computed using the same ITU expression in Eq. (1)). $R_{0.01}$ is rendered in 19 mm/h bins. Surprisingly, the map produced from the ITU method show no ">N" regions in Africa at all. Other notable differences show up in the circled areas on the rain zone map, such as coastal Colombia, southwest India, northeast Bangladesh and northeast India, and much of Southeast Asia. These circled areas are a crude approximation of Zipser *et al.*'s maps ([31], notably Fig. 7a) of the most frequent locations for intense convective storms; these locations have been part of meteorological "local knowledge" for many years ([31], p.1066), so it is no surprise that the hand-produced rain zone map from 1994 should be similar. While not all rain is convective, much of the most intense rain is; as 0.01% exceedance probability corresponds to only ~52 min per year, we should be suspicious of any $R_{0.01}$ map that fails to capture Zipser's intense convective storm locations. On the other hand, the rain zone map sets out very large regions as having homogeneous heavy rain due to the sparsity of the (largely pre-satellite) data used to produce it.

The numeric column in Figure 1 shows the estimated population affected by ">N" conditions (i.e., total population of all pixels in the study area where $R_{0.01} > 95$) according to each method. The 4x difference between ITU and TMPA obviously indicates a fair amount of disagreement in terms of impact; in particular, the ITU figure suggests that high rain rates affect hardly any of the 5 billion people in the study area.

The disagreements evident in Figure 1 provide motivation to construct a new rainfall climatology, one that results in a spatial distribution of $R_{0.01}$ that qualitatively resembles the rain zone map (broadly matching our qualitative knowledge of the location of rainy places) but is more quantitatively consistent with other climatologies (reflecting 20+ years of observational data).

IV. CONSTRUCTION OF A NEW CLIMATOLOGY

The main motivation to merge satellite data with rain gauge data is to build on their respective strengths: wide coverage vs. direct measurement. As we have seen, different climatologies are based on different combinations of data sources and how they are merged. The goal here is to construct a climatology with minimal processing that enables us to provide better estimates of $R_p$.

*A. The TRMM mission and precipitation radar*

The main dataset used here comes from the TRMM precipitation radar [32] (hereafter, *PR*). The TRMM satellite operated in an inclined, non-Sun-synchronous, low-Earth orbit from 1997-2015; while there are many other precipitation measurement missions, the TRMM PR dataset is unique in its combination of active (radar) measurement, total global coverage of the (sub-)tropics (36°S-36°N, 180°W-180°E), and measurement duration (1998-2014).

The TRMM 2A25 data product consists of pixel-wise rain statistics along with corresponding geolocations [8]. The TRMM PR scanned across its swath at ~1.6 Hz, producing observations in multiple range bins at 49 angles ("rays") for each of ~8000 scans over each ~91 min. orbit. As such, each ray corresponds to a "stack" of pixels along a line from the spacecraft to the Earth's surface. The pixel geolocations do not form a regular grid on the Earth's surface since they are defined by the combination of the spacecraft's position, its attitude, and the current ray angle. The pixel size also varied over time as a function of spacecraft altitude [21].

---

[1] In [26], the authors interpolate TRMM 3B42 (TMPA, at 1/4°) to estimate $M_t$ and TRMM 3A25 (radar observations, at 1/2°) to estimate $P_0$. These are plugged into the ITU model above to estimate $R_p$. When we say that we are using TCC, GPCC and WCLIM to estimate $R_p$, we are using the climatology in question to estimate $M_t$ and applying the method of [26] to estimate $P_0$.

[2] Though free-space optical (FSO) vendors often claim that rain is not an important attenuation source for FSO links, 100 mm/h is still expected to cause rain attenuation losses on the order of 10-20 dB/km (see, e.g., [29], [30]), which can halve the vendor-recommended range. As such, *heavy* rain turns out to be material for FSO as well.

Unlike rain gauges, weather radars do not measure rain volume directly. The TRMM PR was necessarily calibrated in many different ways: rain statistics were calibrated against a NASA ground validation network [33], radar surface reference data was incrementally collected at a fine resolution [34], etc. Nevertheless, radar-specific factors such as non-uniformity (e.g., rain in only part of a large pixel), path attenuation (e.g., measuring at multiple points along a rainy path), sensitivity (e.g., inability to detect very light rain), and interference (e.g., ground clutter) naturally results in some systematic differences relative to what a rain gauge network would report.

### B. Previous PR-derived products

Naturally, there are many research data products derived from TRMM PR. For example, the TRMM team has itself produced a range of PR-derived data products that attempt to mitigate biases, whether by binning observations into large spatiotemporal bins (TRMM 3A25), merging TRMM PR data with external data sources (TRMM 3B42 / TMPA [24]), or using ensembles of all of the TRMM sensors (TCC [21]).

Most relevant to this paper are the few global (e.g., [35], [36]) and many regional (e.g., [37]–[40]) data products that have attempted to improve upon the TRMM team's products by binning TRMM PR data at a finer resolution and, in some cases, merging the PR data with regional rain gauge data. The smallest grid size used in these products is generally ~1/24° (vs. 1/4° for TMPA and 1/2° for TCC and 3A25) as ~5 km is comparable to the underlying PR pixel size and smaller bins would result in spatial under-coverage.

Inherent in the binning approach is a loss of spatial registration. That is, binning is fast and can be done with desktop GIS software, but it treats each observation as a point, aggregating observations by their centers. The smoothing caused by spatiotemporal aggregation is desired, but the aliasing caused by the fixed grid is not.

### C. New TRMM-derived estimates of $M_t$ and $P_0$

While spatial resolution is necessarily limited by the PR pixel size, we need not accept additional loss due to spatial (mis)registration. Here, we describe the construction of a sub-pixel PR/gauge product (SPRG) that places the PR pixels on a sub-resolution grid to improve registration.

*1) Initial estimates of $M_t$ and $P_0$*

We first extract all TRMM 2A25 observations classified as "rain certain" with near-surface rain rate (NSRR) >0 mm/h, out of a total of ~44 billion observations. We render these sequentially onto a 1/120° grid on the WGS84 Earth ellipsoid. That is, the geolocation (lat/lon) is used as the center of a circular footprint (an approximation – the PR phased array sweeps a Gaussian beam pattern [32] along a narrow swath) that is 4-5 km wide depending on current TRMM altitude [21]. Since a 1/120° pixel can be observed by multiple PR pixels in a short window (from adjacent rays or scans), we retain the highest NSRR for that pixel measured during this window.

TABLE II. COMPARATIVE SUMMARY STATISTICS.
(Best in **boldface**, worst <u>underlined</u>.)

| short name | $\bar{M}_t$ (mm) | | rel. error, GHCN-D (%) | | |
|---|---|---|---|---|---|
| | land | pop>0 | mean | sd | rms |
| TCC | 910 | 1079 | 0.2 | 27.2 | 27.2 |
| WCLIM | 937 | 1123 | 4.2 | 16.8 | **17.4** |
| *SPRG* | *964* | *1139* | *1.7* | *24.8* | *24.8* |
| GPCC | 976 | 1167 | 4.3 | 17.1 | 17.7 |
| ITU | 999 | 1183 | 9.8 | 29.4 | <u>31.0</u> |
| TMPA | 1014 | 1199 | 9.4 | 21.9 | 23.8 |

The result is, for each 1/120° pixel, a total number of observations; a total number of "rain certain" observations (with NSRR >0 mm/h); and the summed NSRR values. The estimated conditional rain rate for each 1/120° pixel is just the mean of its NSRR values and its estimated $P_0$ is just the ratio of its "rain certain" and total observations. The initial estimate of $M_t$ is the product of conditional rain rate (in mm/y) and $P_0$.

*2) Calibration / merging*

As just mentioned, estimates derived from satellite data tend to have systematic biases relative to reference data. Typical adjustment methods [41] range from mean-field bias correction (MBC) to various forms of kriging. Simple methods such as MBC are still common (including in TMPA [24] and ITU [42]) due to their robustness.[3] These adjustments are often weighted based on covariates, such as pixel-wise error statistics (e.g., [45]) or elevation (e.g., [46]).[4] Here, we have chosen to adjust the 2A25-derived $M_t$ grid with the GPCC $M_t$ grid (smoothed by passing a 121x121 uniform filter over the GPCC data, interpolated to the same resolution; the window size motivated in part by stationarity arguments in [51]). However, because we found GPCC to be unreliable in orographically complex situations, the GPCC adjustment is inverse-weighted by elevation variability (log of interquartile range within the same window, as computed from mean GMTED elevation at the same resolution [52]; the window size in this case motivated by discussions of orographic scale in [25]).

The final SPRG $M_t$ and $P_0$ grids are smoothed using a 21x21 Gaussian filter to address artifacts resulting from the size/spacing of the PR footprints (this window roughly covers a pixel's ray/scan neighbors).

---

[3] Kriging would be the obvious choice in geostatistics, but it is common for regional studies to find mixed results for kriging variants, since rain gauge networks are often very sparse – e.g., [43], [44].

[4] Topographic features are used in regional climatologies such as PRISM [47] where local knowledge is available, but less commonly used for global climatologies (CRU CL [48] and WorldClim [22] being notable exceptions). The relationship between elevation and rainfall is complex, and straightforward use of elevation in 3D interpolation or as a covariate remains controversial [25], [49]. Here, we are using elevation as a proxy measure of error, more along the lines of the inverse error weighting used in TMPA [24] and GPCP [50]. Station density could be a more direct error measure for the gauge data [45] but we do not have the GPCC station density at an appropriate resolution.

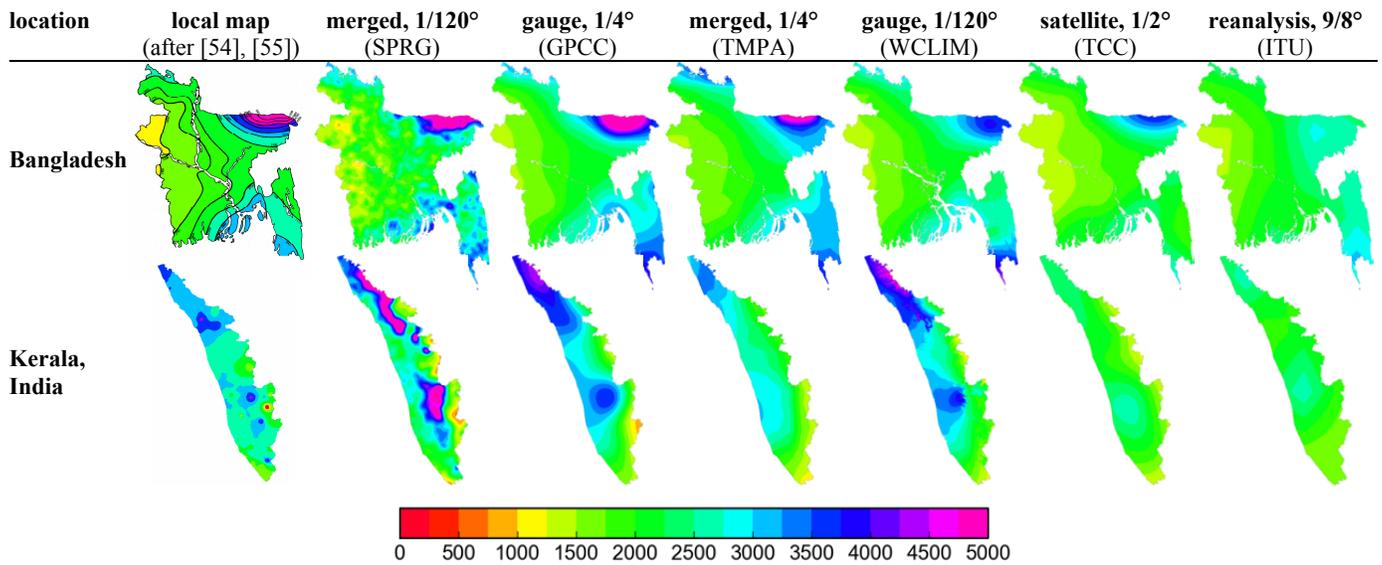

FIGURE 2. ANNUAL TOTAL RAINFALL $M_t$ (MM, 250 MM BINS) FOR TWO SOUTH ASIAN REGIONS AT ~5° SCALE.

### D. Initial plausibility checks of $M_t$

Before proceeding with deeper comparisons, we check whether or not SPRG is physically plausible at various scales.

*Bias consistency:* As in Table I, we use $\bar{M}_t$ (mean $M_t$ taken over all land and populated pixels) as an indicator of relative bias between datasets. As we can see from Table II, SPRG falls squarely in the middle of the reference climatologies (repeated here from Table I); if SPRG has a systematic bias relative to the (unknown) true $\bar{M}_t$, it is not the outlier in this regard.

*Spatial consistency:* Benchmarking spatial variation is more difficult since we do not have access to the WMO's large historical station dataset, which would allow us to check $M_t$ for individual pixels against corresponding station observations. Instead, we check our pixel/station error using an extract of NOAA's smaller GHCN-Daily (GHCN-D) dataset [53]. This extract includes all GHCN-D sites for which there is any subset of 10+ calendar years (from the period 1985-2014) for which each year is 90%+ complete.[5]

Table II shows mean $\mu$, standard deviation (sd) $\sigma$, and root-mean-square (rms) $(\mu^2 + \sigma^2)^{1/2}$ of the relative errors between each site's $M_t$ and its corresponding climatology pixels' $M_t$. The gauge-only climatologies (GPCC, WCLIM) have the lowest bias and variance relative to the GHCN-D station data, with both at <18% rms. (This is to be expected since GHCN data forms the largest single component of GPCC and WCLIM.) All of the satellite (TCC) and merged satellite/gauge (SPRG, ITU, TMPA) products have notably higher variance, with ITU the worst at 31% rms. Again, SPRG is not the outlier, being squarely in the middle.

*Qualitative visual consistency:* None of the climatologies show marked visual differences at global-scale (not shown). It is more instructive to look at examples at a smaller scale. Figure 2 shows two monsoonal regions (Bangladesh and the Indian state of Kerala) at ~5° scale. The maps in the leftmost column are adapted from locally-produced isohyet maps ([54], [55]) which, while not "ground truth," would be expected to reflect local expertise [25].

Bangladesh is a case where the methods largely agree in spatial distribution, disagreeing mainly in magnitude. The climatologies (columns) are ordered left-to-right by the degree of variation (i.e., ITU shows the least overall variation). Bangladesh is relatively simple from an orographic perspective, and all methods agree that most of Bangladesh receives 1500-2500 mm on average. They do disagree considerably about the amount of rainfall in the rising hills in the northwest, northeast and southeast.

Kerala has more orographic complexity, as the Western Ghats (a mountain range) run down its length and close to its coast. Kerala includes the windward side of the Western Ghats (resulting in a strong orographic effect to its west), the Palghat Gap in its center (greatly reducing the orographic effect there), and the leeward side of the Western Ghats (producing a low-rainfall orographic shadow to its east). Far greater differences can be seen for Kerala – the different climatologies show different locations for the regional maxima/minima, as well as very different magnitudes for those maxima/minima. Nevertheless, we can see gross spatial similarities. (While the columns no longer show a straightforward ordering in terms of variation, ITU is again the method with the least overall variation. The differences in variation will be helpful in understanding later results.)

To move on to comparison based directly on $R_p$, we will need an expanded "test set" as the portion of the ITU test set that applies to our study area contains only 34 sites. Such an expanded test set is described in the next section.

---

[5] Applying a quality threshold does make the extract less globally representative. Current data in GHCN-D is overwhelmingly from the USA and Australia, and only 30% (29,242/98,035) of GHCN-D stations are from the remaining 216 countries/territories; the same proportion (3,883/12,318) of the stations in our extract are non-USA/Australia, but only 66 of the other countries/territories are represented.

TABLE III. CHARACTERISTICS OF THE EXPANDED TEST SET.

| site locations | data source | # sites | duration (years) | | | $M_t$ (mm), ITU | | |
|---|---|---|---|---|---|---|---|---|
| | | | min | mean | max | min | mean | max |
| various | ITU test set [9] (study area) | **34** | 1 | **5.4** | 10 | 978 | **1907** | 2975 |
| various, Africa | research literature | 10 | 1 | 2.1 | 3 | 797 | 1459 | 2960 |
| various, S/SE Asia | research literature | 29 | 1 | 3.6 | 10 | 651 | 2473 | 3846 |
| USA & territories | ASOS AWS network [57] | 227 | 5 | 9.4 | 15 | 129 | 1053 | 1632 |
| Australia & territories | BoM AWS network | 43 | 6 | 9.8 | 10 | 405 | 1028 | 1798 |
| Bangladesh | BMD AWS network | 5 | 1 | 1.0 | 1 | 1564 | 1922 | 2757 |
| | | **348** | | **8.5** | | | **1276** | |

TABLE IV. RAIN ZONE COVERAGE STATISTICS FOR ITU AND EXPANDED TEST SETS.

| rain zone | $R_{0.01}$ (mm/h) | land (%, px) | pop > 0 (%, px) | pop (%, pop) | ITU test (%, sites) | expanded (%, sites) |
|---|---|---|---|---|---|---|
| A | 8 | 4.1 | 0.2 | 0.2 | | |
| C | 15 | 11.7 | 2.2 | 2.2 | | |
| D | 19 | 1.8 | 2.7 | 2.1 | | 2.6 |
| E | 22 | 17.5 | 15.1 | 10.3 | | 15.2 |
| F | 28 | 2.7 | 0.1 | 0.0 | | 0.3 |
| H | 32 | 0.0 | 0.1 | 0.2 | | |
| J | 35 | 5.4 | 9.1 | 3.1 | | 0.3 |
| K | 42 | 12.7 | 17.8 | 18.9 | 2.9 | 8.6 |
| M | 63 | 5.7 | 6.3 | 11.0 | 22.9 | 39.1 |
| N | 95 | 21.1 | 29.9 | 38.5 | 22.9 | 21.0 |
| P | 145 | 16.9 | 16.2 | 12.5 | 51.4 | 12.6 |
| Q | 115 | 0.3 | 0.5 | 0.9 | | 0.3 |

## V. CONSTRUCTION OF AN EXPANDED TEST SET

The expanded test set consists of 1-min integration period rain rate statistics from 348 sites (including 34 used in the ITU test set, as described in [9] but restricted to the study area 35°S-35°N), with a mean measurement duration of 8.2 years (vs. 5.4 for ITU) (Table III).

### A. Data sources

73/348 of the site statistics were extracted from databases (34/73) and papers (39/73), and the remainder were created from rain gauge time-series obtained from national meteorological services (275/348). Automated weather station (AWS) rain gauge records were obtained where available at zero/low cost: the USA (and territories), Australia (and territories), and Bangladesh. The tipping-bucket readings were cleaned and converted into 1-min rain rate time-series using the NASA PMM spline interpolation algorithm [56].[6] Isolated physically implausible events (>2"/min) were removed. For each site, the time-series selected was the longest run of 12-month periods for which >90% of records contained valid data; three sites with a large proportion of implausible observations were manually removed.

For consistency with the ITU test set [16], $R_p$ values were only analyzed when the exceedance probability $p$ and measurement duration were such that $R_p$ represented at least 20 observations. (For example, four years of 1-min data were required for $R_{0.001}$ to be considered.)

### B. Categories of diversity

There is no standard metric by which we claim that the expanded test set is "more representative" than the ITU test set, but the following aspects indicate improvement:

*Rainfall diversity:* By including a greater number of low- and mid-rainfall sites, the expanded test set mean $M_t$ ($\bar{M}_t = 1276$ mm, as computed from ITU $M_t$ values at each site) is closer to the overall study area mean ($\bar{M}_t = 1189$ mm, computed from ITU $M_t$ values from all populated land pixels (Table I)). More importantly, the overall spread ($129 \leq M_t \leq 3846$ for the expanded test set, vs. $978 \leq M_t \leq 2975$ for ITU) exercises the ITU model over a wider range of inputs.

*Rain rate diversity:* To check whether the increased range of input $M_t$ is matched by increased rain rate coverage, we can look at the PN.837-1 rain zones as a heuristic stratification by rain rate. The expanded test set touches 9/12 of the land rain zones in the area of study (vs. 4/12 for ITU) (Table IV), with the AWS and non-AWS data each adding two unique zones. The rain zone coverage of the expanded test set is not completely "representative" of the entire study area (as shown in the area, pixel and population statistics in Table IV), being overly weighted toward Zone M instead of Zones K and N, but the expanded test set does clearly represent more rain zones.[7]

---

[6] Tipping-bucket gauges record time-of-tip, not actual instantaneous rain rates. Interpolating time-of-tip to rain rate tends to preserve high rain rates (>1 tips per minute, if integrated at 1-min) but – by design – it also tends to increase $P_0$ and reduce light rain rates.

TABLE V. STANDARD ITU ERROR FIGURES, USING EXPANDED TEST SET.
(Best in **boldface**, worst <u>underlined</u>.)

| short name | 1% rel. error (%) | | | 0.1% rel. error (%) | | | 0.01% rel. error (%) | | | overall rel. error (%) | | |
|---|---|---|---|---|---|---|---|---|---|---|---|---|
| | mean | sd | rms | mean | sd | rms | mean | sd | rms | mean | sd | rms |
| SPRG | -20.5 | 35.4 | 40.9 | 5.4 | 33.1 | 33.5 | -0.1 | 27.5 | **27.5** | -3.1 | 30.9 | **31.0** |
| ITU | -4.6 | 41.9 | 42.1 | -9.2 | 28.5 | **30.0** | -10.0 | 26.6 | 28.5 | -8.5 | 30.9 | 32.1 |
| TCC | -25.4 | 35.5 | <u>43.7</u> | 4.8 | 36.3 | 36.6 | 2.0 | 36.3 | 36.3 | -3.3 | 34.8 | 35.0 |
| WCLIM | -20.7 | 33.4 | 39.3 | 19.6 | 37.8 | 42.6 | 10.3 | 47.7 | 48.7 | 5.6 | 39.7 | 40.1 |
| TMPA | -17.7 | 36.5 | 40.6 | 26.3 | 37.2 | 45.5 | 15.8 | 57.3 | <u>59.4</u> | 10.6 | 44.6 | 45.9 |
| GPCC | -20.0 | 33.2 | **38.8** | 26.3 | 46.7 | <u>53.6</u> | 14.3 | 55.3 | 57.1 | 9.6 | 46.4 | <u>47.3</u> |

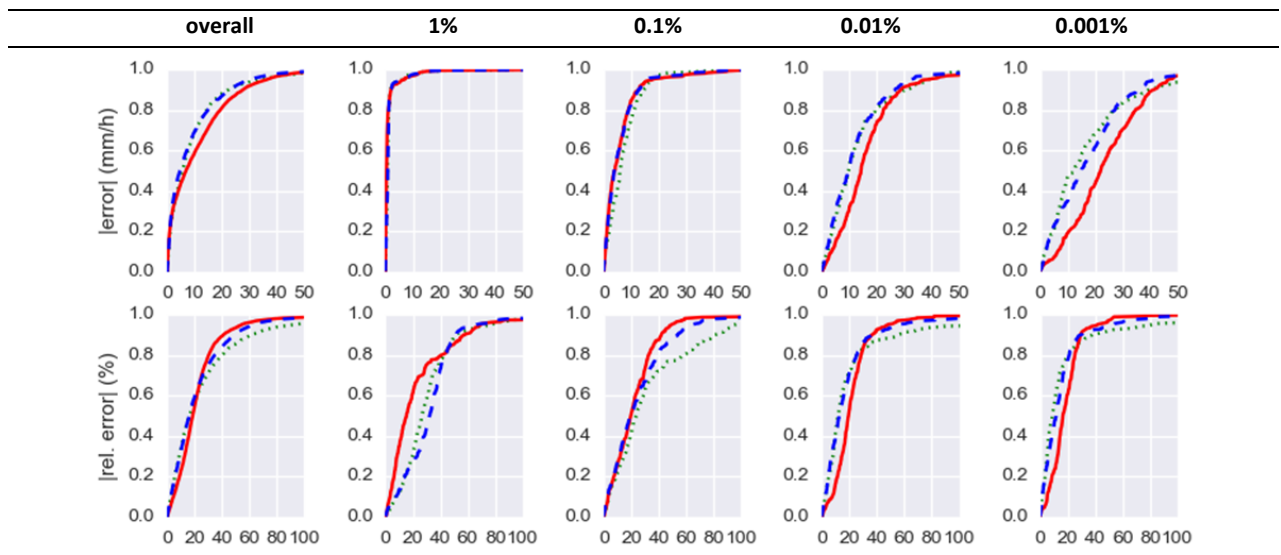

FIGURE 3. CUMULATIVE DISTRIBUTIONS OF REC "ACCURACY."
REC "accuracy" ≡ fraction of sites below the given error threshold [58]; as with ROC curves [59], points toward the upper left indicate superior performance. REC plots correspond to ITU (solid), TMPA (dot) and SPRG (dash).

*Country diversity:* The expanded test set touches 20 countries (vs. 8 for ITU). Since AWS data was available at zero/low cost for only three countries, most of this increase in country diversity comes from the non-AWS data; 84/348 sites are outside of the continental USA and Australia (vs. 28/34 for ITU). Of the 39 non-ITU site statistics, 10 are from sub-Saharan Africa and 29 from South/Southeast Asia.

## VI. COMPARISON

In this section, we compare the performance of each method of estimating $R_p$ using two approaches: (1) conventional error figures at all sites in the expanded test set over a range of exceedance probability values $p$ and (2) by-country identification of rainy areas at $p = 0.01\%$.

### A. Comparing site-wise conventional error figures

There are ITU-recommended ways to compare the effectiveness of propagation-related performance measures. In this subsection, we briefly describe them and then discuss how the estimation methods compare.[8]

*1) Explanation of ITU-R P.311 summary statistics*

The P.837 recommendations were derived using relative error (in %) as the main error figure. As previously mentioned, the Salonen-Baptista rain rate prediction model involves a least-squares fit based on Eq. (1). The fit is evaluated using (signed) relative error $\varepsilon_r = (\hat{r}_{i,p} - r_{i,p})/r_{i,p}$ (as opposed to bias error $\varepsilon_b = \hat{r}_{i,p} - r_{i,p}$ or absolute error $|\varepsilon_b|$), which is evaluated for the model-estimated rain rate $\hat{r}_{i,p}$ and observed rain rate $r_{i,p}$ for each exceedance probability $p$ and for each site $i$ contained in the ITU training set [9]. This use of relative error emphasizes fit quality over the overall (5%-0.001%) range of scenarios, whereas absolute error would emphasize fit quality at the large rain rates corresponding to small $p$.

---
[7] The entire ITU test set represents many of these rain zones, of course, but using sites in well-gauged developed countries in the mid-latitudes.

[8] We do not use evaluation schemes that would involve re-fitting the rain rate estimation model, such as cross-validation (see, e.g., [60]). The ITU training set was hand-selected by experts for subjective representativeness and data quality [9], so selecting many random subsets from a pool of highly variable quality would miss the point.

There are many ways to summarize error figures (MAE, RMSE, etc.). The ITU-R P.311 series [11] recommends the use of a specific set of summary statistics (mean $\mu$, standard deviation (sd) $\sigma$, and rms = $(\mu^2 + \sigma^2)^{1/2}$ ) of the chosen error figure. Reporting all three enables the reader to understand the relative contributions of bias and variance to rms (see, e.g., [60]) and this is what we show in Table V, for all six methods at three sample $p$ values as well as the aggregate over all $p$ values. [9]

*2) Explanation of plots*

Figure 3 shows cumulative distribution of "accuracy" vs. error threshold for ITU, SPRG and TMPA. This visualizes fit quality as classification accuracy; instances are considered correctly classified if the error falls within the given threshold value (the $x$-axis). The cdf plots of this non-decreasing "accuracy" function are referred to as REC curves [58] by way of analogy to conventional ROC curves [59]. For each exceedance probability $p$, we show cdf plots for both absolute error $|\varepsilon_b|$ and absolute relative error $|\varepsilon_r|$.

*3) Comparison of conventional error figures*

From Table V and Figure 3, we observe the following:

First, SPRG and ITU are very similar in terms of rms error when averaged over all exceedance probabilities (and indeed at the individual exceedance probabilities as well) (Table V). The overall REC-"accuracy" plots also show modest differences between ITU (solid) and SPRG (dash) (Figure 3). From Table V, both appear (superficially) to be better than the four other methods; in these examples, ITU or SPRG is usually the best (bold) and never the worst (underlined). Of particular note is that they do better at 0.01%.

Second, variance is very high throughout Table V, and as a result no method does better than ~30% rms error on the expanded test set. This is consistent with previous evaluations (see, e.g., [9], [26], [27]), which often justify this level of error by noting that inter-annual rainfall variation is also ~30% and that most site statistics are derived from much less than the 30-50 years of data that are typically used to compute rainfall "normals" (e.g., [61]).

Third, these high variances (which are obviously the dominant contributor to rms error throughout Table V) obscure bias differences that are themselves fairly large. As can be seen from the REC-"accuracy" plots of both absolute error $|\varepsilon_b|$ and absolute relative error $|\varepsilon_r|$ for ITU and SPRG (Figure 3), there are large $|\varepsilon_b|$ differences at small $p$ (high rain rates) and considerable $|\varepsilon_r|$ differences across all $p$. Of particular note is that ITU does much better at $|\varepsilon_r|$ for large $p$ (e.g., $p$ =1%) where the corresponding $|\varepsilon_b|$ is small.

Because the rms errors are so high, the most that we can conclude is that, qualitatively, SPRG and ITU are more consistent in reaching the ~30% level than the other methods. (Similar results have been achieved in other recent attempts to improve upon ITU, e.g., [62].) The fact that all methods have weaknesses (resulting in high variance at varying $p$) makes it difficult to cleanly separate them (it would be useful, for example, if ITU or SPRG were always best), but the lack of robustness of GPCC, WCLIM and TMPA does suggest that there is no clear reason to use them for our application (estimating $R_p$).

TABLE VI. "CLASSIFICATION" CONFUSION MATRIX AND DERIVED MEASURES (EXPANDED TEST SET COUNTRIES ONLY).
(Best in **boldface**, worst underlined.)

| short name | by site | | | | by country | | | |
|---|---|---|---|---|---|---|---|---|
| | confusion matrix | | acc. [0,1] | MCC [-1,1] | confusion matrix | | acc. [0,1] | MCC [-1,1] |
| | F | T | | | F | T | | |
| **ITU** F | 266 | 2 | 0.81 | 0.34 | F 5 | 2 | 0.45 | 0.02 |
| T | 65 | 15 | | | T 9 | 4 | | |
| **Zone** F | 262 | 6 | **0.82** | **0.43** | F 3 | 4 | 0.60 | 0.12 |
| T | 55 | 25 | | | T 4 | 9 | | |
| **SPRG** F | 259 | 9 | 0.81 | 0.38 | F 6 | 1 | **0.70** | **0.45** |
| T | 56 | 24 | | | T 5 | 8 | | |

*B. Comparing classification accuracy*

Table VI summarizes performance on the expanded test set, showing confusion matrices of classification by-site (sites in the expanded test set that are correctly identified as ">N" or not) and by-country (countries represented in the expanded test set that contain at least one site that is correctly classified as ">N" or not). The confusion matrices are further summarized with accuracy and correlation (here, the Matthews correlation coefficient, which accounts for class imbalance – see, e.g., [63]).

*By site:* The classification performance of the three methods is not clearly distinguishable. Each does better than the ~77% baseline accuracy implied by the class proportions (i.e., a classifier that guesses F on every input would achieve $80/348 \approx 77\%$ accuracy).

*By country:* In contrast to by-site classification, by-country classification accuracy not only shows a clear ranking (ITU, Zone, SPRG) but also highlights that only SPRG does better than the $13/20 \approx 65\%$ baseline accuracy implied by the class proportions. MCC also shows a consistent ranking. On both measures, ITU is the worst due to its high false negative rate; this false negative rate highlights that the low bias shown for ITU in the $|\varepsilon_b|$ and $|\varepsilon_r|$ REC-"accuracy" curves (the $p$=0.01% column of Figure 3) is spread across many countries, not just a few.

On all classification measures, then, ITU is clearly the method that is least effective at identifying ">N" regions. This is most clear for by-country classification.

---

[9] Note that we do not apply the additional weighting for site measurement duration and intra-latitude-band variance used in [9] – the many-fold increase in longer duration, higher-latitude sites (Table III) already has the same effect.

TABLE VII. THE SUBSET OF STUDY AREA COUNTRIES FOR WHICH ANY
METHOD IDENTIFIED >10M PEOPLE IN ">N" CONDITIONS.
Clusters (boldface) can be seen where SPRG aligns with **ITU against Zone**;
with **Zone against ITU**; and with **neither**.

|  | any site statistics >N? | total in study area (pop, M) | pop >N (pop, M) | | |
|---|---|---|---|---|---|
|  |  |  | Zone | SPRG | ITU |
| (global) |  | 5,136 | **690** | **678** | 263 |
| **India** | - | 1,205 | 0 | **130** | 1 |
| **Bangladesh** | + | 161 | 12 | **126** | 12 |
| **Indonesia** | + | 249 | 249 | **122** | **111** |
| Philippines | - | 104 | 34 | 28 | 43 |
| **Nigeria** | + | 170 | **86** | **27** | 0 |
| **China** | + | 900 | 0 | **27** | 0 |
| Vietnam |  | 92 | 0 | **24** | **23** |
| Burma |  | 55 | 35 | 23 | 19 |
| **Brazil** | + | 199 | 52 | **17** | **9** |
| Malaysia | + | 29 | 29 | 15 | 18 |
| **Sri Lanka** | + | 21 | 0 | **10** | 0 |
| **Nepal** |  | 30 | 0 | **10** | 0 |
| **Taiwan** |  | 23 | 0 | **8** | 0 |
| **Mexico** |  | 115 | 0 | **8** | 0 |
| **Thailand** | + | 67 | 26 | **7** | **11** |
| Colombia | - | 45 | **11** | 7 | 0 |
| **Cameroon** | + | 20 | **14** | 5 | 0 |
| **DR Congo** |  | 74 | **11** | 5 | 0 |
| **Madagascar** |  | 22 | **15** | 4 | 0 |
| **Côte d'Ivoire** |  | 22 | **12** | 2 | 0 |
| **Ghana** |  | 25 | **18** | 2 | 0 |

We now consider our actual impact metric: the pixel-wise regional performance of Zone, ITU and SPRG, as summarized by estimated population affected by ">N" conditions. Again, our unit of analysis will be countries; the affected pixel or population totals cannot be verified, but countries are very meaningful from a market analysis and project execution (e.g., regulatory) perspective.

Table VII illustrates the agreement between methods, for all countries in the study area for which any of the methods indicated that there were >10M people. The numbers in the rightmost columns show the total population within the part of each country within the study area that is affected. The clusters (boldface) represent cases where SPRG aligns with ITU against Zone, with Zone against ITU, or is an outlier. (As a check, the table also indicates where we have site statistics in the expanded test set that confirm ">N" conditions or not.) With regard to some of the main patterns from the Zipser map, we can see that:

- SPRG agrees with Zone (over ITU) that multiple countries in sub-Saharan Africa do contain ">N" regions, and this is confirmed by site data in (e.g.) Nigeria and Cameroon.
- SPRG agrees with ITU (over Zone) that Brazil and several Southeast Asia markets (notably Indonesia) are likely less consistently ">N" than suggested by Zone.
- SPRG agrees with neither that multiple countries in South Asia and East Asia do contain ">N" regions, and this is confirmed by site data in (e.g.) Bangladesh, Sri Lanka and China.

To summarize Table VII in a different way: simply using Zone (PN.837-1) or ITU (P.837-5/6) has some significant network planning implications – they *often* disagree, the disagreement is *not* always in the same direction, and the answer (as indicated by the expanded test set) is that there *are* populations affected by ">N" conditions in these countries more often than not. As such, it highlights that there is business decision risk to "business as usual" with regard to rain rate prediction.

VII. CONCLUSIONS

The current ITU-R P.837 method is a remarkable achievement, producing stable rms error even when tested against a much larger test set. However, predictors that are robust in terms of rms error are often "lukewarm," i.e., have reduced dispersion around the mean, and we saw anecdotal evidence of this in Figure 2. Since the highest rain rates correspond to regions where $M_t > 2000$ mm and the overall mean $M_t$ over land is only ~1000 mm (for all climatologies considered here), we might expect a low bias in exactly the (rainy) places that concern us most. This is borne out by Figure 3 and Table VI, and likely explains curious results such as ITU predicting no ">N" regions in Africa (even though ITU's overall rms error at $p$=0.01% is similar to its error at any other exceedance probability).

*A. Implications*

By increasing test set size and diversity, we have shown that the predictions of ">N" conditions made by the ITU method appear to be anomalous and that other methods can produce more accurate results. So far, the best balance of rms error and classification error has been from SPRG; though residual plots (not shown) do suggest that better results can still be obtained.

Based on this outcome, it seems reasonable to assume that the SPRG population impact numbers are more reliable than those from Zone (PN.837-1) or ITU (P.837-5/6). Certainly, the fact that SPRG's true-positive sites are spread over twice as many countries vs. ITU (8/20 vs. 4/20) (Table VI) indicates improved generality. Some hypotheses implied by this assumption (even if we do not yet have a test set that allows us to validate them fully) would be:

- The total affected population is closer to the 690M estimated by Zone than to the 263M estimated by ITU, but the affected population is less concentrated than suggested by Zone.
- Large regions with ~1B population (sub-Saharan Africa, India) that have no ">N" population according to ITU do appear to be affected.
- Southeast Asia is likely to be a better market for high-capacity wireless backhaul than suggested by Zone.

- *All* of the countries listed in Table VII (as well as others not shown) contain *some* ">N" population according to SPRG, so high-capacity backhaul projects must consider local conditions (rain statistics) carefully. This is trivial advice in general – ITU P.837 gives a version of it as well – but this paper has illustrated its geographic and demographic impact in specific.

Reduced reliance on the old, uneven rain zone maps should build confidence the reliability of high-capacity wireless in (sub-)tropical areas.

### B. Potential improvements

Beyond generic needs for better models, more data, and more error/sensitivity analysis, there are several ways in which this work could be developed:

*Effects of very light and very heavy rainfall:* Table V shows that, for relative error, the variance component of rms error dominates the bias component for all rainfall datasets at all exceedance probabilities. Residual plots (not shown) suggest that outliers in low rain rate regions (< 1000 mm/h) are problematic.

*Dataset refinement:* Certain foibles of the TRMM PR are not addressed here. For example, there are TRMM PR calibration issues around inland/coastal water [34] as well as issues around snow and ground feature resolution at high altitudes [32]. Similarly, adjustment for within-pixel spatial heterogeneity (see, e.g., [64]) could be made (beyond those already incorporated in the TRMM 2A25 product, e.g., [8]). Finally, we have not rigorously quantified the sampling bias (see, e.g., [65]) remaining in the full 17-year record.

While the results can always be improved and made more rigorous, we believe that a clear case has been made here for (1) greater investigation of bias/variance issues during future revision of P.837, (2) the need for renewed consideration of heavy rain rate in deploying millimeter-wave links in former Zone P/Q areas (such as much of West Africa), and (3) the greater applicability of millimeter-wave technology in former Zone P areas (such as much of Southeast Asia).

SUPPLEMENTARY INFORMATION I: REFERENCES FOR SITE STATISTICS OBTAINED FROM RESEARCH DATABASES AND PAPERS

The table below provides references for site statistics extracted from the literature.
- Where data was not provided in tabular form, cdf/ccdf figures were traced and interpolated to standard probability values using log-linear interpolation.
- Rows marked with "*" replace shorter-duration site statistics in DBSG3 provided by the same authors.
- Some of the pre-1994 site statistics appeared in CCIR (pre-DBSG3) databases.

| site | locality | ctry. | yrs. | refs. | site | locality | ctry. | yrs. | refs. |
|---|---|---|---|---|---|---|---|---|---|
| Annaburroo | Annaburroo | AUS | 3 | [1], [2] | NARL | Gadanki | IND | 6 | [8] |
| Bathurst | Bathurst | AUS | 2 | [1], [2] | VSSC | Thiruvananthampuram | IND | 6 | [8] |
| Boa Vista | Boa Vista | BRA | 1 | [1], [2] | Intelsat | Nairobi | KEN | 2 | [9] |
| LBA | Ji Parana | BRA | 1 | [1], [2] | NUOL | Vientiane | LAO | 1 | [10], [11] |
| Macapa | Macapa | BRA | 1 | [1], [2] | Paksong | Paksong | LAO | 1 | [10], [11] |
| Mosquiero | Mosquiero | BRA | 2 | [1], [2] | UM | Moratuwa | LKA | 1 | [12] |
| Natal | Natal | BRA | 2 | [1], [2] | Kwajalein | Kwajalein | MHL | 8 | [1], [2] |
| Ponta Das Lagos | Ponta Das Lagos | BRA | 2 | [1], [2] | DID | Bintulu | MYS | 5 | [13] |
| Santarem | Santarem | BRA | 1 | [1], [2] | SMV | Bota | MYS | 1 | [14] |
| Sao Paulo | Sao Paulo | BRA | 3 | [1], [2] | MEASAT | Bukit Jalil | MYS | 1 | [15], [16] |
| Chongqing | Chongqing | CHN | 10 | [1], [2] | FTA | Cyberjaya | MYS | 1 | [1], [2] |
| Dongxing | Dongxing | CHN | 10 | [1], [2] | MMU | Cyberjaya | MYS | 1 | [1], [2] |
| Fuzhou | Fuzhou | CHN | 10 | [1], [2] | DID | Dalas | MYS | 5 | [13] |
| Guangzhou | Guangzhou | CHN | 10 | [1], [2] | UTM-JB | Johor Bahru | MYS | 9 | [17]–[19] |
| Guilin | Guilin | CHN | 10 | [1], [2] | DID | Kiansam | MYS | 5 | [13] |
| Hangzhou | Hangzhou | CHN | 10 | [1], [2] | UTM-KL | Kuala Lumpur | MYS | 3 | [1], [2], [20] |
| Nanchang | Nanchang | CHN | 10 | [1], [2] | DID | Kuching | MYS | 5 | [21] |
| Nangjing | Nangjing | CHN | 10 | [1], [2] | DID | Kuhara | MYS | 5 | [13] |
| Wuhan | Wuhan | CHN | 10 | [1], [2] | DID | Miri | MYS | 5 | [13] |
| Yichun | Yichun | CHN | 10 | [1], [2] | USM-KCP | Perak | MYS | 2 | [22] |
| Intelsat | Douala | CMR | 2 | [3] | DID | Stapang | MYS | 5 | [13] |
| Ayura | Ayura | COL | 4 | [1], [2] | FUTA | Akure | NGA | 3 | [23] |
| Convento | Convento | COL | 4 | [1], [2] | OAU | Ile-Ife | NGA | 2 | [24] |
| Cucaracho | Cucaracho | COL | 4 | [1], [2] | UNILAG | Lagos | NGA | 2 | [25] |
| Gerona | Gerona | COL | 4 | [1], [2] | LAUTECH | Ogbomoso | NGA | 2 | [26] |
| Girardota | Girardota | COL | 4 | [1], [2] | OSU | Osogbo | NGA | 1 | [27] |
| Manantiales | Manantiales | COL | 4 | [1], [2] | Covenant | Ota | NGA | 1 | [28] |
| Pedregal | Pedregal | COL | 4 | [1], [2] | UNIYOLA | Yola | NGA | 1 | [29] |
| San Cristobal | San Cristobal | COL | 4 | [1], [2] | UNITECH* | Lae | PNG | 5 | [4] |
| USP* | Suva | FJI | 5 | [4] | Si Racha | Si Racha | THA | 3 | [1], [2] |
| Cibonong | Cibonong | IDN | 2 | [5] | Klongyai | Klongyai | THA | 1 | [30] |
| Padang | Padang | IDN | 2 | [5] | Houston | Houston | USA | 8 | [1], [2] |
| Putussibau | Putussibau | IDN | 2 | [5] | Jacksonville | Jacksonville | USA | 8 | [1], [2] |
| ITS | Surabaya | IDN | 3 | [6] | Tampa | Tampa | USA | 5 | [1], [2] |
| Tanahmerah | Tanahmerah | IDN | 2 | [5] | Durban | Durban | ZAR | 1 | [31] |
| GNDU | Amritsar | IND | 1 | [7] | | | | | |